\documentclass{appolb}
\usepackage{graphicx}
\usepackage{amssymb}
\usepackage{hyperref}

\begin{document}
\title{Study of optical and gamma-ray long-term variability in blazars
\thanks{Presented at Krakow Epiphany 2022 international conference}%
}
\author{Gopal Bhatta
\address{Institute of Nuclear Physics, Polish Academy of Sciences PAN\\
PL-31342 Krakow, Poland}
}
\maketitle
\begin{abstract}
Blazars, a subset of powerful active galactic nuclei, feature relativistic jets that shine in a broadband electromagnetic radiation, e. g. from radio to TeV emission. Here I present the results of the studies that explore gamma-ray and optical variability properties of a sample of gamma-ray bright sources Several methods of time-series analyses are performed on the decade-long optical and Fermi/LAT observations. The main results are as follows: The sources are found highly variable in both the bands, and the gamma-ray power spectral density is found to be consistent with flicker noise suggesting long-memory processes at work. While comparing two emission, not only the overall optical and the $\gamma$-ray emission are highly correlated but also both the observation distributions exhibit heavy tailed log-normal distribution and linear RMS-flux relation. In addition, in some of the sources indications of quasi-periodic oscillation were revealed with similar characteristic timescales in both the bands. We discuss the results in light of current blazar models with relativistic shocks propagating down the jet viewed close to the line of sight.
\end{abstract}
  
\section{Introduction}
Active galactic nuclei (AGN) are one of the most luminous sources in the Universe. The sources are powered by the accretion on to the supermassive black holes at the center of the distant galaxies. The objects can be broadly classified in two main groups: radio-loud AGN and radio-quiet AGN. While a large fraction of AGN do not show prominent radio emission, 
 about 10\% of them output a significant amount of radio emission \cite{urry1995unified}. The emission can be linked to the presence of kpc/Mpc scale radio jets as seen in the VLBI images. The bipolar jets are often found to remain highly collimated over large distances and mostly contain plasma matter traveling at the nearly at the speed of the light. In a small sub-group of the radio-loud AGN, the jet is aligned to the line of sight at $\lesssim 6^o$ \cite{Jorstad2017}, such that the relativistic beaming effects dominate lending  the sources some of their extreme observational properties such as high luminosity, TeV emission and  rapid variability.

The broadband emission from blazars can be characterized by a double-hump feature in the frequency-flux plane. It is widely accepted that the low frequency feature of the spectrum, lies between radio and X-ray,  arises due to the synchrotron emission by the energetic charged particle moving in the jet magnetic field; whereas the high frequency feature, that lie between UV and gamma-rays, probably results due to the inverse-Compton scattering of the low energy photons by high energy particles. In the synchrotron self-Compton (SSC) leptonic models, the same population of electrons up-scatters the co-spatial synchrotron photons. But in the external Compton (EC) models the low energy seed photon could be contributed by accretion disk, dusty torus and broad-line regions (BLR). Blazars can be further sub-classified into two groups: the more powerful flat spectrum radio quasar (FSRQ) that show emission line over the continuum emission and the less powerful BL Lacs showing weak or no presence of emission lines. The sample of TeV blazars is mostly dominated by BL Lacs.

Blazars display multi-timescales variability across a broad range of electromagnetic spectrum (see e. g. \cite{Bhatta2013,Bhatta2016,Bhatta2018,Mohorian2022}). The basic character of variability appears to be stochastic in nature such that the statistical properties of the variability can be fairly represented by power-law models in the frequency domain. In addition, the sources light curves also show occasional rapid flares and quasi-periodic oscillations \cite{Bhatta2019}

As multi-wavelength (MWL) variability is defining properties of AGN, variability studies are important tool to probe the energetics of the central engines of AGN.
In this proceedings, I discuss the recent results of the studies focused on the time series analysis of a sample of gamma-ray bright blazars. The decade-long gamma-ray observations from Fermi/LAT and the optical observations from four different observatories were analyzed and the results in the two bands were compared.

\begin{table*}
\center
\caption{\label{tab:table1}Sample sources for study of gamma-ray and optical variability of blazars}
\begin{tabular}{|r|l|l|l|l|}
\hline
Source name & 3FGL name & Source class & Red-shift \\
\hline
\textbf{	3C 66A 	}&	J0222.6+4301 	&	BL Lac 	&	0.444	\\
\textbf{	AO 0235+164 	}&	J0238.6+1636	&	BL Lac 	&	0.94	\\
	PKS 0454-234 	&	3FGLJ0457.0-2324	&	BL Lac 	&	1.003	\\
\textbf{	S5 0716+714 	}&	J0721.9+7120 	&	BL Lac 	&	0.3	\\
\textbf{	Mrk 421 	}&	3FGLJ1104.4+3812 	&	BL Lac 	&	0.03	\\
 TON 0599 	&	J1159.5+2914 	&	BL Lac 	&	0.7247	\\
	ON +325 	&	J1217.8+3007 	&	BL Lac 	&	0.131	\\
	W Comae 	&	J1221.4+2814	&	BL Lac 	&	0.102	\\
	4C +21.35 	&	3FGLJ1224.9+2122 	&	FSRQ 	&	0.432	\\
\textbf{	3C 273 	}&	J1229.1+0202 	&	FSRQ 	&	0.158	\\
\textbf{	3C 279 	}&	J1256.1-0547 	&	FSRQ 	&	0.536	\\
\textbf{	PKS 1424-418 	}&	3FGLJ1427.9-4206 	&	 FSRQ	&	1.522	\\
	PKS 1502+106 	&	3FGLJ1504.4+1029	&	FSRQ 	&	1.84	\\
	4C+38.41 	&	J1635.2+3809 	&	FSRQ 	&	1.813	\\
\textbf{	Mrk 501 	}&	J1653.9+3945 	&	 BL Lac 	&	0.0334	\\
	1ES 1959+65 	&	J2000.0+6509 	&	BL Lac 	&	0.048	\\
\textbf{	PKS 2155-304 	}&	J2158.8-3013 	&	BL Lac 	&	0.116	\\
\textbf{	BL Lac 	}&	J2202.7+4217 	&	BL Lac 	&	0.068	\\
\textbf{	CTA 102 	}&	J2232.5+1143 	&	FSRQ 	&	1.037	\\
\textbf{	3C 454.3  	}&	J2254.0+1608 	&	FSRQ 	&	0.859	\\

\hline
	\end{tabular}
	\end{table*}

\section{Observations and Data}
Gamma-ray variability of blazars was investigated using Fermi/LAT observations spanning nearly 10 year from 2008-2018 in the spectral range 100 MeV to 300 GeV.  The sample source along with their Fermi/LAT 4-year Source Catalog (3FGL) name, source classification and red-shift are listed in Table \ref{tab:table1}.  The data processing was performed using the standard procedure of likelihood analysis of the unbinned analysis and weekly binned source light curves were produced. \footnote{\url{https://fermi.gsfc.nasa.gov/ssc/data/analysis/scitools/likelihood_tutorial.html}} (For details see \cite{bhatta2020nature})

That optical data for the sources marked in the fold fonts in Table \ref{tab:table1} were gathered from  four ground based observatories: SMARTS\footnote{\url{
http://www.astro.yale.edu/smarts/glast/home.php}}, Steward observatory\footnote{\url{http://james.as.arizona.edu/~psmith/Fermi/}}, Catalina\footnote{\url{http://nesssi.cacr.caltech.edu/DataRelease/}} and AAVSO\footnote{\url{https://www.aavso.org/}}. The data were complied to create a dense observations constructing 10 year-long optical light curves (For details see \cite{Bhatta2021}).

\section{Analysis and Results}
The decade-long optical and gamma-ray observations were analyzed using several methods of time-series analysis. In particular, the observed variability was quantified by computing fractional variability of the source light curves \cite{Vaughan2003,galaxies6010002}. The large fractional variability values are the indications that blazars exhibit remarkable variability on diverse time-scales. Also to compare the variability in the optical and the gamma-ray band, the results show that  the variability amplitude are larger  in the gamma-ray.

Power spectral density analysis of the gamma-ray light curves was performed by computing the discrete Fourier periodogram of the source light curves. The effects of unevenly sample, discrete observations and finite observation lengths were corrected by the use of a large number mock light curves that were simulated by the Monte Carlo method described in \cite{timmer}. It is found that the power spectral density of the gamma-ray observations are consistent with single power law model with spectral index $\sim$ 1, often known as long-memory flicker noise. The result implies a strong dik-jet connection in the sense that the jet modulations are somehow coupled to the disk modulations. The optical PSD is however steeper, closer to the Brownian motion, than that of the gamma-ray PSD slope index.

The flux distribution of the gamma-ray and optical flux of the blazar was explored by fitting the flux histograms with normal and log-normal probability distribution function (PDF). The analysis revealed that flux distribution can be represented by log-normal PDF \cite{bhatta2020nature}.  Similar results were obtained in a study of optical variability of blazars using decade-long optical observations \cite{Bhatta2021}. In addition, the source light curves in both the bands were divided into several segments and relation between the mean and RMS  of the segments were analyzed. It was found that majority of the sample sources displayed a linear dependence between the RMS and the flux, so called linear RMS-flux relation.

In addition to the PSD and PDF, the light curves were studied to look for possible quasi-periodic oscillations (QPO) in the  light curves. For the purpose the gamma-ray observations were analyzed using the methods  Lomb-Scargle periodogram  \cite{scargle1982,lomb1976} and Z-transformed Weighted Wavelet methods \cite{Grant1996}.  The results showed that the light curves of the sources Mrk 501, Mrk 421, S5 0716+714, PKS 1424-418, ON +325  and PKS 2155-314 revealed year timescale QPOs which were found to be significant over the red-noise inherent in blazar light curves \cite{bhatta2020nature,Bhatta2019}. Moreover,  in the sources Mrk 501, Mrk 421, S5 0716+714 and PKS 2155-314 the observed QPOs were found to have similar characteristic timescales (See Table 4 in \cite{Bhatta2021}). 

Furthermore, a cross-correlation study between the optical and gamma-ray light curves was conducted using z-transformed discrete correlation function (ZDCF) \cite{Alexander2013}, a method frequently used in the discretely sampled data sets with gaps. Moreover, significance of the ZDCF peaks against any spurious correlation arising due to red-noise is computed. The results suggest that both optical and gamma-ray emission from the sources highly correlate with each other within the timescales of a few weeks (See Table 5 in \cite{Bhatta2021}).

\section{Discussion}
A study of long-term variability properties of gamma-ray bright blazar was performed using the gamma-ray observations from the Fermi/LAT space telescopes and the optical observation. In the standard model of blazars the observed variability can be mainly ascribed to the processes occurring at accretion disk and jets of AGN. In the disk scenario, the magneto-hydrodynamics instabilities arising from modulations in the accretion rates, viscosity parameter and the magnetic field can lead to the observed variability. Similarly in jets related scenarios some of the widely discussed models are based on particle acceleration due to shock waves and/or due to turbulence wide spread in the jets, and the subsequent cooling of the particles via dissipative processes such as synchrotron and inverse-Compton emission \cite{Spada2001,marscher2008inner,27rs,marscher2013turbulent}. In such scenarios, variable injection rates could also produce the optical variability \cite{Bhatta2013}. Moreover, strong correlation between the optical and the gamma-ray emission suggests existence of co-spatial particle that contribute to both the emission. In the case of magnetized jets, magnetic reconnection could be a dominant particle accelerating mechanism, and several mini-jets could form within the main jets such that it could be source of TeV emission \cite{giannios2009fast}.  The observed log-normal PDF of blazar flux could be indication of such processes \cite{Biteau2012}. Power-law PSD of slopes of unity in particular describe the long-memory processes at work such that the flux changes in the accretion disk can propagate along the jets and somehow affect the variable processes along the jets. The fact that the source flux distribution are consistent with the log-normal PDF indicates the processes being multiplicatively coupled to result in the total observed variable emission, as opposed to linear combination of additive processes such as shot noises\cite{uttley2005non}. Moreover, the linear dependence between the flux and the RMS, which are more relevant to the galactic X-ray binary systems, are often explained in terms of the propagation of viscosity modulations at the disk \cite{Lyubarskii1997}. That fact that we observe similar trend in the blazar sample sources, where the disk emission is completely swamped by the jet emission, indicates a strong coupling between the jet and disk. 

The peaks in the periodogram that appear in both the bands on similar timescales imply plausibility of MWL QPOs  with a common origin. Such QPO can arise either from the accretion disk and the jets. Some of the widely discussed blazar QPO models include binary supermassive black holes, passage of the emission region along helical jets, Lense-Thirring precession, and periodic jet angle modulations(readers are directed to \cite{Bhatta2016,Bhatta2018,Bhatta2019} for detailed discussion on the topic)

\subsubsection*{Acknowledgment}I acknowledge financial support by the Narodowe Centrum Nauki (NCN) grant UMO-2017/26/D/ST9/01178.

\bibliographystyle{IEEEtran}
\bibliography{ref}
\end{document}